# Scalable Traffic Predictive Analysis using GPU in Big Data

**Dalyapraz Dauletbak, Junghoon Heo, Sooyoung Kim, Yeon Pyo Kim and Jongwook Woo**
Luddy School of Informatics, Indiana University, US
Softzen Co. Ltd., Republic of Korea
Department of Information Systems, California State University Los Angeles, US
[e-mail: dmanato@iu.edu, {jhheo, sykim, ypkim}@softzen.co.kr, jwoo5@exchange.calstatela.edu]
*Corresponding author: Jongwook Woo

***Abstract***

The paper adopts parallel computing systems for predictive analysis in both CPU and GPU leveraging Spark Big Data platform. The traffic dataset is adopted to predict the traffic jams in Los Angeles County. It is collected from a popular platform in the USA for tracking information on the road using the device information and reports shared by the users. Large-scale traffic data set can be stored and processed using both GPU and CPU in this Scalable Big Data systems. The major contribution of this paper is to improve the performance of machine learning in distributed parallel computing systems with GPU to predict the traffic congestion. We show that the parallel computing can be achieve using both GPU and CPU with the existing Apache Spark platform. Our method can be applicable to other large scale datasets in different domains. The process modeling, as well as results, are interpreted using computing time and metrics: AUC, Precision and Recall. It should help the traffic management in Smart City.

## 1. Introduction

GPU supports parallel computing with deep learning libraries. Big Data solution provides distributed file and computing systems. We can achieve better high performance computing by integrating GPUs to Big Data platform. It can reduce the gap between the big data and deep learning communities as well because the deep learning models can read large scale data more efficiently and easily from the distributed file systems of Big Data platform.

Traffic data collection and analysis is significant for Smart City. US Governments turn to Advanced Traffic Management Systems in order to solve traffic congestions and adopt new transport management plans and resources [2].

Juniper Research finds that smart traffic management systems could save cities US$277 billion by 2025 through reducing emissions and congestion [14]. INRIX estimates that traffic congestion cost U.S. commuters $305 billion in 2017 due to wasted fuel, lost time and the increased cost of transporting goods through congested areas [15]. As governments started adopting smart cities' concept for the past decades, traffic prediction attracts more attention.

Our paper presents traffic jam prediction using Spark and Rapids, which is Big Data machine learning engine and GPU parallel computing libraries, respectively. The paper is composed of Section 2. Related Work, Section 3. Big Data and GPU, Section 4. Dataset and Specifications, Section 5. Prediction with Machine Learning, and Section 6. Conclusion.

## 2. Related Work

There is a growing interest in traffic prediction

(Support note) This research was supported as a Brain Pool Program of 2020 by The Ministry of Science and ICT and National Research Foundation of Korea and a research credits program from Google Cloud Platform.

systems to support traffic operators in city's decision-making tasks.

Waze is an app for those who are interested and willing to connect with it for better community. As a crowd sourcing, the users of Waze can exchange data to drive easier with more information and to make data-driven infrastructure decisions and increase the efficiency of incident response [5]. One of the works that is based on traffic data of Waze is available in the form of slides from Summit on Data-Smart Government at Harvard [6]. This study focuses on collaboration of Waze and Louisville City and points out major insights from such partnership. The outcome of this work is analysis of data in the form of animated maps and Excel tables of hot spot traffic [7,8].

Another study was conducted in New Haven County, Connecticut. In this research GPS data set was gathered from MapMyRun traffic website and further processed and analyzed using R [9]. The author used sampled small data set for analysis, whereas we present a framework with bigger data sets. Also, this work concentrates on clustering the hot areas of traffic, however, our work gives prediction of jams using classification model with Big Data platform utilizing GPU.

Dalya et al. explained the flow of big data files management and further prediction of traffic jams using machine learning with Hadoop Spark Big Data platform [16]. But, in this paper, we leverage Big Data platform utilizing GPU to get much better performance for traffic prediction.

## 3. Big Data and GPU

Apache Spark has been one of popular solutions for Big Data. It supports in-memory processing as a distributed parallel computing systems with machine learning libraries. It is integrated into Hadoop Big Data systems as a computing engine. Hadoop cluster is composed of HDFS file systems, computing engines with MapReduce and Spark, and YARN resource management. YARN helps the platform linearly scalable.

### 3.1 Gap in Big Data and Deep Learning

The traditional data science develops machine learning models in Python and R for the small dataset. It has the data size of up to Mega-Bytes and generates memory issues when to process Giga-Bytes of data set. Figure 1 shows the gap between the traditional and Big Data.

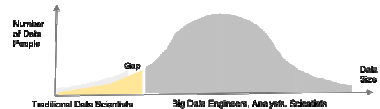

**Fig. 1.** A gap between the traditional Data Scientists / Deep Learning and the professionals in Big Data

Legacy data projects needs more Big Data Engineers, Analysts, and Scientists while the data grows exponentially. Even the deep learning system with a single server has the similar issue when to read and process massive dataset, greater than Giga-bytes.

### 3.2 GPU in Big Data

GPU chip with Multi-cores accelerates the development of the deep learning applications with various Deep Learning algorithms. Deep Learning libraries such as Tensorflow and Keras efficiently use Multi-cores for parallel computing to achieve high performance. And, Andrew Ng at Stanford University shows that Deep Learning models are more scalable and accurate than legacy Machine Learning while data grows larger. Big Data community has been working on leveraging the Big Data platforms with GPUs to read massive dataset stored in HDFS and to utilize both legacy machine learning and deep learning models.

The RAPIDS suite is an open source software libraries to execute machine learning analytics in GPU utilizing NVIDIA CUDA [11]. NVIDIA CUDA is a parallel computing architecture supporting parallel operations. NVDIA has created Rapids for Spark 3.0, which drastically improves the performance of ETL, data engineering, data analysis, and data prediction [12]. Thus, Spark 3.0 in GPU supports deep learning and legacy machine learning as shown in Figure 2. It also resolves the issue of the gap presented in Figure 1 by transferring large scale dataset for GPU to process as illustrated in Figure 2.

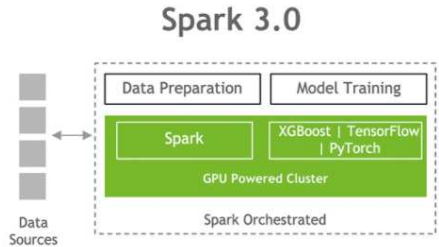

**Fig. 2.** Apache Spark 3.0 in GPU

We can implement Big Data predictive analysis with both legacy machine leaning and deep learning algorithms in Spark and Rapids. That is, we can leverage Big Data cluster with multiple nodes that have both CPUs and GPU. The cluster can achieve high performance with parallel computing operations in GPU chips and distributed parallel computing in CPUs.

Classifications are commonly used to detect Ad Click and credit card Fraud. Decision Trees is one of the algorithm for classification [11]. In the paper, three machine learning algorithms are compared for classifying and detecting the traffic jams: Random Forests, Gradient Booting Tree, and XGBoosting.

Random Forest is ensembles of decision trees by combining many decision trees with being expressed as a set of de-correlated decision trees. Thus, it reduces the risk of overfitting. The example of Random Forest can be a data set that contains different random values and their class. Then the data set is divided into a lot of subsets with random values and random classes. After the division, the algorithm decides and allocates different classes to each of the independent forests. Similarly, Gradient Boosting Tree uses Decision trees as a group of machine learning algorithms. It also combines many weak learning models together to create a strong predictive model.

XGBoost is a distributed gradient boosting library also called GBDT (Gradient Boosting Decision Tree) and GBM (Gradient Boosting Machine). It provides parallel tree boosting, which allows parallel distributed computing possible in Hadoop/Spark cluster [13]. In the paper, we use XGBoostClassifier in Spark for classification.

## 4. Dataset and Specifications

The traffic dataset was provided by Information Technology Agency of Los Angeles City Department for study purposes and consisted of 5,858 JSON files covering information reported by app users (accidents, jams, road closure etc.) and information captured from users' devices (location, speed, time deviation from original route). Since this database is not publicly open and data is shared upon request only, we were authorized to use a portion of the data only. The dataset is of the size 1.8 GB and covers nine days (Dec 31, 2017 – Jan 8, 2018).

The data has two major files: alerts (information reported by users) and jams (information captured by user's device). Total number of rows (event records) for alerts and jams are 2,170,694 and 16,058,236 rows respectively. The same data processes can be applied to much bigger dataset (as large as 70GB+ annually) as Hadoop Spark Big Data systems is linearly scalable.

The Table 1 shows the hardware specification for GCP (Google Cloud Platform) cluster.

| Spark Cluster | 2 worker nodes (CPU) | 2 GPUs |
|---|---|---|
|  | n1-highmem-32 | nvidia-tesla-t4 |
| **Cores** | 32 | 48 |
| **Memory** | 208 GB | 32 GB |

**Table 1.** H/W Specification

The table 2 lists the attributes and metadata of alerts after cleaning the data:

| location_x | X-coordinate of location |
|---|---|
| location_y | Y-coordinate of location |
| street | Street name |
| city | City name |
| country | US |
| road_type | Road type |
| report_description | Small text describing the traffic event written by user |
| type | Type of reported traffic even: road_closed, jam, accident, hazard |
| pub_date | UTC Time of the publication |
| | of traffic report |
| date_pst | Pacific Time of the publication of traffic report |
| month | Month number of the publication (1-12) |
| day | Day of the publication (1-31) |
| hour | Hour of the publication (0-23) |
| min | Minute of the publication (0-59) |
| sec | Second of the publication (0-59) |
| weekday | Day of the week of the publication (Monday - Sunday) |

**Table 2.** Alerts attributes

Table 3 lists the additional attributes and metadata filtered for jams, which is generated passively from device's GPS:

| | |
|---|---|
| level | jam level, where 1 – almost no jam and 5 – standstill jam |
| speed | driver's captured speed in mph |
| length | length of the traffic ahead in the route of user in meters |
| delay | time deviation from the original time in seconds |

**Table 3.** Jams attributes

## 5. Prediction with Machine Learning

### 5.1 Machine Learning Flow

We aim to predict the traffic jam with classification model.

Prior to model training we chose to split dataset into 75% of training set and 25% of model performance testing set. After several iterations of model training/testing and by calculating the weight of the columns, we excluded columns that have no value for traffic jams prediction. We used *TrainSplitValidation* and evaluation set of *xgbClassifier*, which helps build general model.

There are several metrics to validate performance of the multiclass classification model as follows: *AUC (Area Under ROC Curve)* – area of total records classified correctly; *Precision* – ratio of correctly identified records as positive out of total records identified as positive; *Recall* – ratio of correctly identified records as positive out of total actual positives [15].

Our model is built with 2 worker nodes and 2 GPUs shown in Table 1. The number of executor cores are set to 4 and *spark.task.cpus* to 1, so it runs 4 concurrent tasks per executor. XGBoost's NUM_WORKERS and *nthreads* are set for 2 and 1, respectively.

In the range of 5 traffic levels presented in Table 3, we regard that the traffic jam *label* is true when *level* is greater than 2. Table 4 shows the confusion matrix of XGBoost in which GPU accelerates the computation to build the traffic prediction model.

| | Predicted 0 | Predicted 1 |
|---|---|---|
| **Actual 0** | 2,259,865 | 0 |
| **Actual 1** | 0 | 4,398,279 |

**Table 4.** Confusion Matrix of XGBoost Model

We weight more with Recall and AUC for the accuracy. Recall become higher when false negative (FN) is smaller. FN means in traffic prediction is when the model predicts no traffic jam but it actually has the traffic jam. AUC generally states the percentage of accurate prediction.

| | RF | GBT | XGBoost |
|---|---|---|---|
| **AUC** | 86.3% | 89.6% | 100% |
| **Precision** | 0.890 | 0.922 | 1.0 |
| **Recall** | 0.956 | 0.947 | **1.0** |
| **Computing Time** | 1 hrs 8 min 53 sec | 3 hrs 55 min 23 sec | **21 sec** |

**Table 5.** Accuracy Measurement

In Table 4 with traffic labeling, XGBoost has all 100% for AUC and Recall with the parameter pairs of maximum depth and leaves: (*max_depth*: 5, *max_leaves*: 256). Its computing time with GPU is amazingly short comparing to the traditional machine learning models, RF and GBT.

## 6. Conclusion

We present Big Data platform and architecture that provides distributed file systems with parallel computing of CPUs utilizing GPUs. It leverages the Big Data platform for storing and analyzing giga-bytes of data set in parallel computing with both CPUs and GPUs. Furthermore, the architecture is linearly scalable with possibly more data set.

We compare three algorithms to predict traffics jams in Los Angeles: Random Forrest, Gradient Booting Tree, and XGBoosting. Our experimental result shows that XGBoosting has the highest performance with the perfect accuracy for predicting traffic jam. Moreover, it is accelerated by GPU and achieves highest processing time in seconds.

Further work can be done with more servers, bigger dataset, and other classification models using deep learning algorithms in order to find more insights and create a data driven conclusions on LA County traffic situation by using this framework to reach the goal of smart city.

## References

[1] M. Heiskala, J. Jokinen, and M. Tinnilä, "Crowdsensing-based transportation services — An analysis from business model and sustainability viewpoints," Research in Transportation Business & Management, Vol 18, pp. 38-48, 2016.
[2] J. Barbaresso, G. Cordahi, D. Garcia et al., "USDOT's Intelligent Transportation Systems (ITS) ITS Strategic Plan 2015-2019," 2014.
[3] "Integrated Corridor Management," Intelligent Transportation Systems - Integrated Corridor Management, www.its.dot.gov/research_archives/icms/. Accessed April 14, 2019.
[4] J. Kestelyn, "Real-Time Data Visualization and Machine Learning for London Traffic Analysis," Google Cloud, 2016, cloud.google.com/blog/products/gcp/real-time-data-visualization-and-machine-learning-for-london-traffic-analysis. Accessed April 14, 2019.
[5] "Connected Citizens by Waze," Waze, www.waze.com/ccp. Accessed April 14, 2019.
[6] M. Schnuerle, "Louisville and Waze: Applying Mobility Data in Cities," Harvard Civic Analytics Network Summit on Data-Smart Government, 2017.
[7] Louisville Metro. "Thunder Jams, 2017 Traffic Delays." CARTO, louisvillemetro-ms.carto.com/builder/d98732d0-1f6a-4db2-9f8a-e58026bf0d39/embed. Accessed April 14, 2019.
[8] Louisville Metro. "Pothole Animation." CARTO, cdolabs-admin.carto.com/builder/a80f62bf-98e1-4591-8354-acfa8e51a8de/embed. Accessed April 14, 2019.
[9] E. Necula, "Analyzing Traffic Patterns on Street Segments Based on GPS Data Using R," Transportation Research Procedia, Vol. 10, pp. 276–285, 2015.
[10] J. Woo and Y. Xu, "Market Basket Analysis Algorithm with Map/Reduce of Cloud Computing," in Proc. of International Conference on Parallel and Distributed Processing Techniques and Applications (PDPTA), Las Vegas. 2011.
[11] "RAPIDS, Open GPU Data Science" Web Documentation, https://rapids.ai/. Accessed April 23, 2021.
[12] "GPU Accelerated Apache Spark", Web Documentation, https://www.nvidia.com/en-us/deep-learning-ai/solutions/data-science/apache-spark-3/. Accessed April 23, 2021.
[13] "XGBoost Documentation", Web Documentation, https://xgboost.readthedocs.io/en/latest/. Accessed April 23, 2021.
[14] "Smart Traffic Management could save cities", Web Documentation, https://cities-today.com/smart-traffic-management-could-save-cities-us277-billion-by-2025/. Accessed April 7, 2021.
[15] "Advanced traffic management next big thing smart cities", Web Documentation, https://www.greenbiz.com/article/advanced-traffic-management-next-big-thing-smart-cities Accessed April 7, 2021.
[16] D. Dauletbak and J. Woo, "Big Data Analysis and Prediction of Traffic in Los Angeles," KSII Transactions on Internet and Information Systems, vol. 14, no. 2, pp. 841-854, 2020. DOI: 10.3837/tiis.2020.02.021